**Title:**

Integration of Scanning Probe Microscope with High-Performance Computing: fixed-policy and reward-driven workflows implementation

**Authors:**

Yu Liu[1*], Utkarsh Pratiush[1], Jason Bemis[2], Roger Proksch[2], Reece Emery[1], Philip D. Rack[1], Yu-Chen Liu[4], Jan-Chi Yang[4], Stanislav Udovenko[5], Susan Trolier-McKinstry[5], and Sergei V. Kalinin[1,3*]

[1] Department of Materials Science and Engineering, University of Tennessee, Knoxville, Tennessee, 37996 USA
[2] Oxford Instruments Asylum Research, Santa Barbara, California 93117, USA
[3] Physical Sciences Division, Pacific Northwest National Laboratory, Richland, WA 99354
[4] Department of Physics, National Cheng Kung University, Tainan, 70101, Taiwan
[5] Materials Science and Engineering Department, Materials Research Institute, the Pennsylvania State University, University Park, PA 16802, USA

* Corresponding author: yliu206@utk.edu, sergei2@utk.edu

**Abstract**

The rapid development of computation power and machine learning algorithms has paved the way for automating scientific discovery with a scanning probe microscope (SPM). The key elements towards operationalization of automated SPM are the interface to enable SPM control from Python codes, availability of high computing power, and development of workflows for scientific discovery. Here we build a Python interface library that enables controlling an SPM from either a local computer or a remote high-performance computer (HPC), which satisfies the high computation power need of machine learning algorithms in autonomous workflows. We further introduce a general platform to abstract the operations of SPM in scientific discovery into fixed-policy or reward-driven workflows. Our work provides a full infrastructure to build automated SPM workflows for both routine operations and autonomous scientific discovery with machine learning.

**Introduction:**

The extensive application of scanning probe microscopy (SPM) has opened the doors to explore and modify the nanoworld. Compared to other materials characterization tools, SPM offers a desktop footprint, low cost, and versatility in operating in multiple environments [1, 2]. It provides a wide range of functional imaging capabilities, extending from basic topographic imaging [3-7] to probing of electronic [4, 8], magnetic [3, 9, 10], mechanical [3, 4], biological [7, 11-15], and chemical [16, 17] properties. Furthermore, SPM supports multiple spectroscopy techniques in a variety of imaging modes, enabling comprehensive understanding and manipulation of materials at the nanoscale [18-20].

However, since the early days of SPM, the principles of operation remained the same – image scans and point spectroscopy at locations chosen by operators. To systematically explore the variability of physical properties across the sample surface, hyperspectral measurements in which spectroscopy is performed on a grid was introduced. However, full-grid spectroscopy measurements are time-consuming and risky in terms of both probe and sample damage. In most scenarios, only spectroscopy around specific structural features such as step edges, domain boundaries, grain boundaries, strained regions, and other defects is of interest. One specific example is that in topological materials, only I-V spectroscopy around step edges is needed to distinguish topological edge states [21-24]. On the other hand, sometimes researchers are interested in the discovery of nanoscale structural elements that manifest specific functionalities that can be detected by spectroscopy measurements. Examples include searching for structural features that give rise to higher transition temperatures in superconductors [25], lower current onset voltage in semiconductors [26-28] and larger piezoresponse in ferroelectric materials [29, 30].

With the development of computer vision and machine learning (ML) [31-34], attempts have been made to integrate machine learning methods and scanning probe microscopy. These were preponderantly realized in the form of workflows in which execution of the codes is driven by immediately available targets via fixed policies. For example, this can include the use of the deep convolutional networks or simpler image analysis tools for the identification of the *a priori* known object of interest such as atoms in scanning tunneling microscopy [35], identification of single DNA molecules [36], spectroscopy of grain boundaries [37], and ferroelectric domain walls [38-42]. More complex examples entail inverse workflows, in which the goal is to discover the structural features that maximize the desired aspect of the spectral response [29].

The number of examples of ML integration into active SPM workflows has been growing rapidly over the last 2-3 years [18, 43-47]. Combined with the rapidly growing number of possible ML algorithms for computer vision and image segmentation, optimization, and other tasks, this suggests tremendous opportunities for much more complex developments. However, SPM generates imaging and spectral data at a high speed, which correspondingly requires substantive computational power to handle the data with advanced algorithms, necessitating the use of the external computational resources, ideally a high-performance cluster computing in real time.

Here, we present a general framework for integrating SPM with high-performance computing (HPC). We introduce a Python interface that mimics the actions of human operators, allowing the SPM to be controlled from either a local computer or a remote cloud server. This interface allows users to build their own workflows for routine operations. With this interface, autonomous workflows based on fixed policy and reward-driven algorithms were implemented. Several exemplar scenarios that benefit from these autonomous workflows are then described, including optimization of domain writing voltage in the piezoelectric force microscopy (PFM) mode, combinatorial library exploration on a grid, performing spectroscopy only on selected features, and discovery of structural features based on spectral features of interest.

**I. Typical tasks in automated microscopy**

As a first step, some of the most common tasks that emerge in the context of the scanning probe microscopy experiment are described.

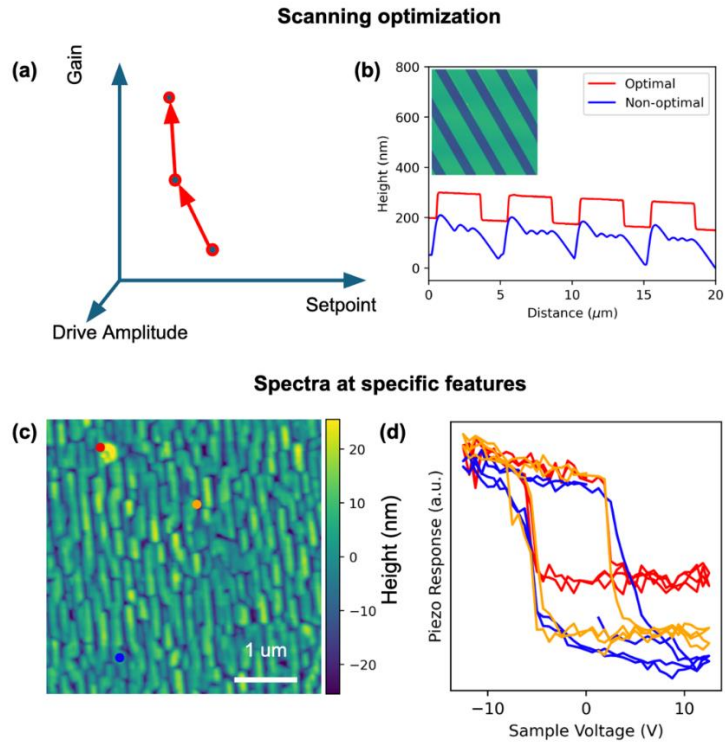

**Figure 1.** Tasks for autonomous SPM workflows. **(a).** To optimize an AC scan, it usually requires fine tuning of multiple scanning parameters. **(b).** Example scan lines with suboptimal and optimal scanning parameters. The sample is Asylum Research (AR) Height calibration grating sample. The inset shows the topography map in the same area as the line scans. **(c).** Topography map taken on ferroelectric $PbTiO_3$. The colored dots mark three different structural features, including a higher grain (red), a normal grain (blue), and a grain boundary (orange). **(d).** Hysteresis loop of piezoresponse measured in (c) at the locations with corresponding color codes. The spectra are correlated with the local structural features around the measurement location. A common task necessitates identification of certain structural features in a fast topography map to select locations for time consuming measurements.

**Imaging optimization:** Imaging optimization for different scanning modes is an essential automated workflow in SPM [3, 4]. Conventionally, users take a lot of time and effort to fine tune the scanning parameters for a specific material and imaging mode. For example, to optimize the scanning parameters for the AC tapping mode, the most widely used imaging mode in SPM, there are many factors to consider, as shown in Figure 1(a). A high-resolution scan in the tapping mode usually requires a slow scan speed, a small oscillation amplitude of the probe and gentle setpoint. However, the roughness of the surface can limit the smallest oscillation amplitude of the probe. Similarly, a larger integral gain (I Gain) in the PID loop ensures the probe responds faster to the change in the height, but too large of a gain will lead to instabilities and positional oscillations [48-50]. These factors can lead to poor images and probe or sample damage. In particular, even a single high-force interaction with the sample can irreversibly damage the probe, requiring replacement, or worse, leading to poor data. Practically, users must balance the imaging quality, acquisition time and safety of the probe. The complexity of these various choices can lead to poor reproducibility, even with experienced users.

**Spectroscopy on selected objects – including both spectroscopies and high-resolution images:** Another typical task for automated microscopy is to acquire time-consuming spectroscopic or high-resolution imaging measurements only on selected features. These features can be step edges, specific types of defects, certain domain boundaries or grain boundaries, etc. as shown in Figure 1(c). The conventional approach of acquiring spectroscopy data on a grid is unnecessarily time consuming and unsafe to the probe as most of the measurement time is wasted on the background. Thus, it's advantageous to take spectroscopy data only on selected features. This concept can be applied to any situations where fast measurements are used to locate the features of interest, and then time-consuming measurements like spectroscopy, high-resolution scans, and other advanced imaging modes are performed only around these features.

**Discovery based on spectral features:** Sometimes, the areas of interest are based on spectral features related to the sample physical properties, rather than structural features. For example, the size of the superconducting gap around the Fermi level in the I-V curve reflects the superconducting transition temperature of the material. Therefore, regions of the sample with large superconducting gap size can then be used to discover the relation between large gap size and underlying structural features. The same concept also applies to many other scenarios: correlating large loop areas in the hysteresis loop spectroscopy with the domain and grain structure in ferroelectric materials or correlating a small switch-on voltage with the structural features in semiconductors. In this type of workflow, the discovery of physics is based on detection of specific spectral features.

**Combinatorial libraries/large sample exploration:** Combinatorial library exploration is another application that can benefit greatly from automated microscopy [51-56]. In a combinatorial library, multiple compositions are produced on the substrate during growth, as shown in Figure 2(a). Thus, continuous variation of composition can be explored on the same substrate, which offers a high-throughput approach to correlating properties with composition. In the example shown in Figure 2(b), the composition of the combinatorial library changes linearly along the horizontal direction of a pseudo-binary composed of $(CrVTaW)_xMo_{1-x}$. As a result, the topography scans taken at three different locations on the library show systematic evolution in the structural features (Figure 2(c-e)), reflecting the underlying crystal phase and structure evolution with composition. Automated workflows ideally can map the composition-property relation without the need for taking time-consuming high-resolution scans and spectroscopy on the full grid. Ultimately, information from different channels like imaging and spectroscopy or different instruments like SPM and X-ray diffraction (XRD) can be unified to provide in-depth knowledge about the combinatorial library.

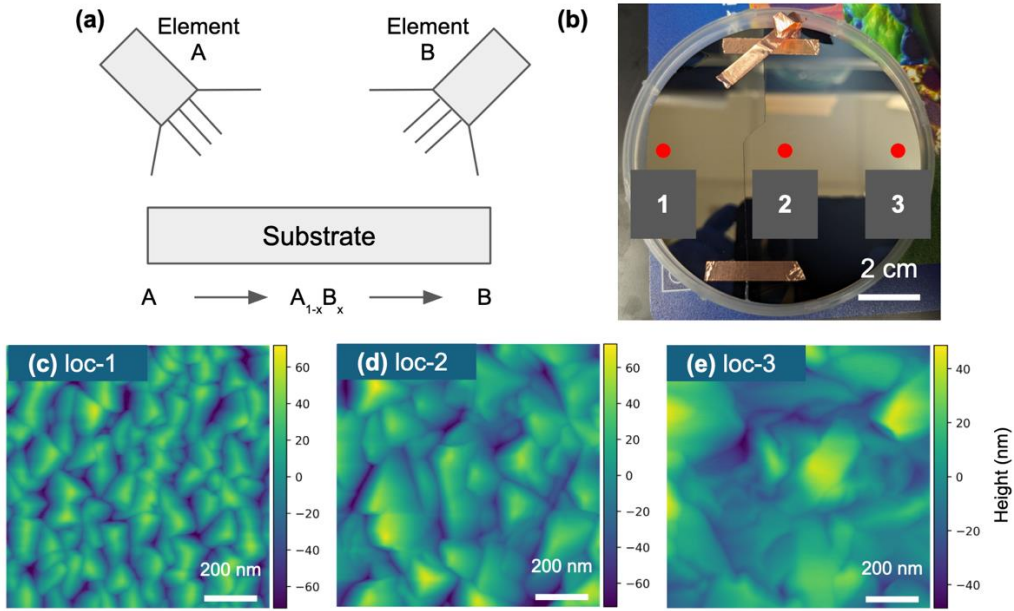

**Figure 2.** Combinatorial library exploration on a grid. **(a).** Schematic illustration of a combinatorial library. The constituents A and B vary along the longitudinal direction of the substrate. The physical properties measured at any location on the library can be directly related to the composition by the location of the measurement. **(b).** A picture of $(CrVTaW)_xMo_{1-x}$ combinatorial library. **(c-e).** Topography maps measured at three locations labeled in (b). At different locations, the structural features change systematically as the composition varies, reflecting the underling phase and crystal structure evolve with the composition.

## II. General framework

We introduce two types of automated workflows–fixed-policy and reward-driven workflows. Generally, a policy is defined as the rule by which machine learning algorithm selects action based on the collected data. Reward is defined as a numerical measure of experimental success derived from the collected data. If the reward is available each step of the process, this is a myopic reward-driven workflow. Otherwise, this is a non-myopic workflow. Note for development of the ML driven workflows, the reward has to be available at least at the end of the experiment. This differentiates the reward form objective, representing the broader goal of the experiment.

Most of the research to date has been based on the fixed-policy workflows. In this case, the nature of the policy is defined before the experiment as a rigid decision-making process. It can be a conditional policy in which an operation will be executed once the conditions are met or a probabilistic one in which decisions are made with certain probability once conditions are met. As an example of workflow with a conditional fixed-policy, after acquisition of a scan image, a computer vision (CV) algorithm like a Canny filter or variational autoencoder (VAE) is employed to detect the location of certain features. Then the workflow will move the probe only to these locations to take spectroscopy measurements. Here, the condition is the presence of specific structural features in the image and the policy is that once a specific feature is detected by CV algorithms, the workflow will move the probe to its location to take spectroscopy.

Reward-driven workflows are more natural to human operators. Here the policies are tuned to maximize the reward. The reward function can be either automatically derived from the collected data, which leads to an optimization problem, or based on human feedback of the collected data, for human-in-the-loop workflows [57, 58]. Examples of reward-driven workflows include tuning the microscope to achieve a better performance. Human operators define a reward function reflecting the scan quality based on the acquired image. Then the workflow will optimize the scan automatically by maximizing the reward function in the parameter space. In reward-driven workflows, human operators focus on creative activities like defining the goal of the experiment and the corresponding reward function. Such repetitive, subjective fine-tuning processes cab be replaced by the workflow and algorithms.

A critical bottleneck in implementing reward-driven workflows is definition of the appropriate reward functions to reflect the goals of the experiment. This requires users to have a deep understanding of the domain knowledge. In addition, in many applications, there are more than one local minima in the reward function. Users need to evaluate the optimization results to make sure they are reasonable. In other applications, the goal of the experiment cannot be defined by a single reward function. Instead, users must consider multiple reward functions, sometimes defined by different types of measurements.

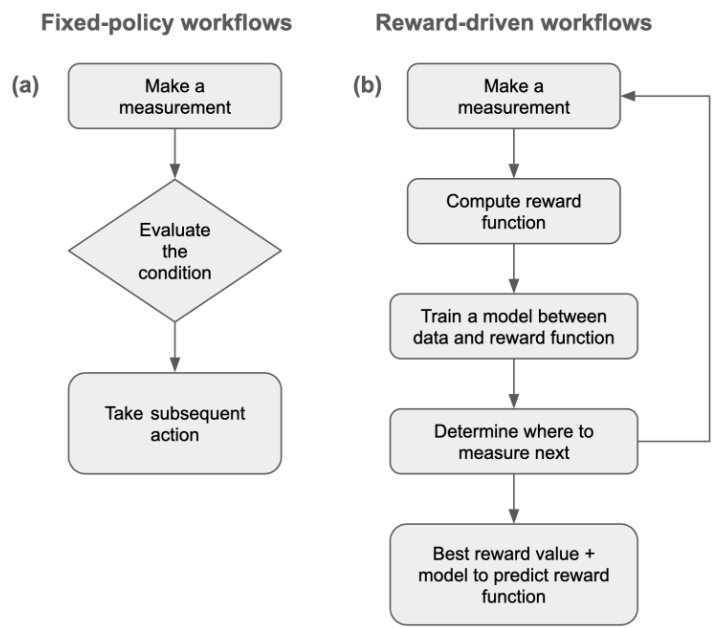

**Figure 3.** Algorithms with fixed-policy and driven by rewards. **(a).** A general diagram of fixed-policy workflow. Here the subsequent action will only be executed when the condition is met, where conditions are based on the acquired data and can be detection of a specific feature or flatness of the scan. **(b).** A general diagram of reward-driven algorithms. Here the workflow is myopic as the reward from the previous measurement will be used to tune the parameter in the next measurement, which is closer to the nature of human operators.

## III. Implementation of the Jupiter-HPC workflows

To enable automated workflows for SPM, we consider the general scheme for the implementation, as shown in Figure 4(a). First, users need to provide the fixed-policy and/or reward functions to define the goal of the experiment. Second, users define the optimization parameter space and institute programmatic control of these parameters. Third, users search parameter space to understand the behavior of the instrument interacting with the sample and identify parametric danger zones. Fourth, users determine appropriate myopic algorithms to optimize the reward functions in the parameter space. Finally, users will deploy the automated workflows on the instrument and evaluate their performance.

To automate operation of the microscope, we have developed an interface enabling Python-based control of the instrument from either local computers or remote supercomputers in Figure 4(b). Control is achieved by Python notebook that encodes sequences of operations into a buffer command file in a hyper-language format. Subsequently, the Python code invokes the SPM controller to execute these operations. To retrieve the status of the instrument, the Python code instructs the SPM controller to write the corresponding parameters in a buffer data file, from which the status can be read. Since communication between code and instrument occurs via buffer files, seamless integration of supercomputers into the workflow is feasible by manipulating these files from the cluster.

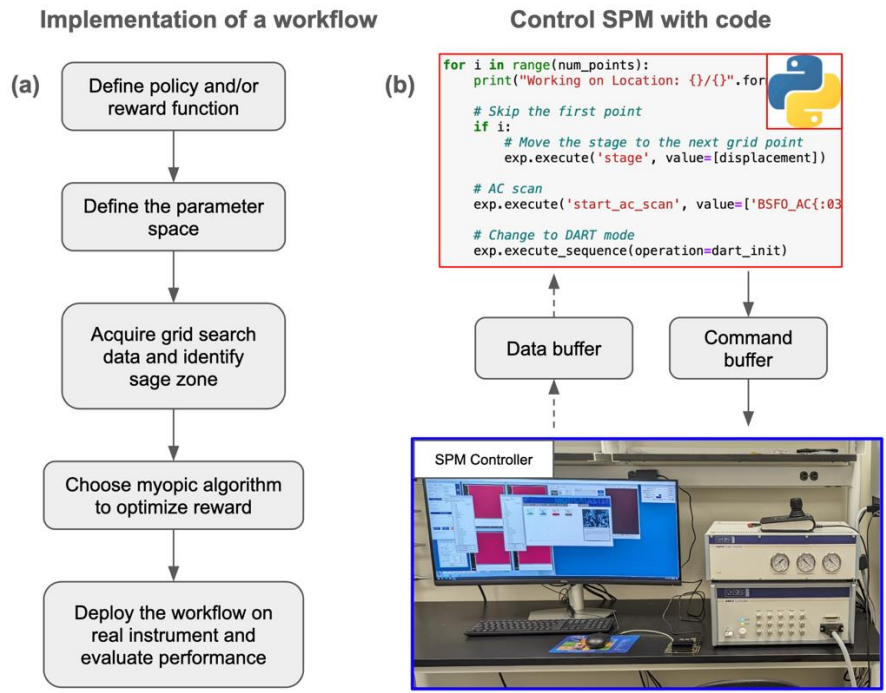

**Figure 4.** Implementation and setup. **(a).** A general scheme of implementing a new workflow. **(b).** The communication between the Python notebook running on a remote supercomputer and the SPM controller is realized through an interface library which consists of a data buffer and a command buffer.

This interface library is designed to emulate the actions of human operators. It offers basic actions as LEGO-style building blocks for users to construct their own workflows to automate repetitive, time-consuming works. The interface library and all the following workflows are tested on a Jupiter SPM by Asylum Research, Oxford Instruments. However, by writing a custom layer that translates the hyper-language commands into a set of controller-specific commands, this interface and all the implemented workflows should be able to run seamlessly on other SPM controllers.

In addition to the normal image and spectroscopy data saved on the disk, the interface library also has access to the intermediate data stored in the controller memory. For example, it can access the scanning trace and retrace lines in real time, which enables tuning the scanning parameters in real time, instead of waiting for completion of the whole image. Another example is that it can initiate and read the tuning data, which facilitates automation of all contact-mode based imaging modes. To summarize, the interface library offers code control for the SPM beyond what a human operator can do. Table 1 lists the delay time for common operations on the Jupiter SPM.

Table 1. Timing of different actions by code from both a local computer (conventional way of control) and a remote supercomputer (ISAAC).

| Actions | Control from local computer | Control from remote server |
| --- | --- | --- |
| Write commands to buffer | 3 ms | 9 ms |
| Execute commands in controller | 70 ms | 75 ms |
| Read the status of SPM | 45 ms | 62 ms |
| Start a scan | 1.5 s | 1.63 s |

## IV. Examples

This section describes four examples of automated workflows. The focus is on definition of the reward functions, how to determine the parameter space and what optimization algorithms can be used. Most importantly, the section also highlights how the interface library makes these automated instrument controls possible.

## IV.a. Gain optimization in PFM

These automated workflows offer an opportunity to study the interplay between the driving voltage and writing voltage in the PFM study of ferroelectric materials. To switch domain orientations in a ferroelectric material, a DC voltage between the probe and sample is required to align the polarization of the domains with the applied electric field. It is also necessary to apply an AC drive voltage to read out the domain orientation. Therefore, we build a workflow to test the minimum DC voltage required to flip ferroelectric domains for different read-out drive voltages. The parameter space in this experiment is hence two-dimensional space of sample voltage and drive voltage.

To define the reward during the experiment, we use the flipping rate as defined in the Figure 5. Movement of the probe along the slow-scan direction was disabled to avoid artifacts. Different combinations of drive voltage and sample voltage are swept in each measurement cycle. At the start of the cycle, the domain structure along the scan line is erased at a large DC voltage of 12.5 V to make sure the starting phase is uniform. After that, a combination of drive voltage and sample voltage is applied to write the domains, followed by a reading cycle at drive voltage =1 V and sample voltage = 0 V. The flipping rate is computed as how many phase pixels have changed from the before and after writing (based on PFM phase data). The reward function at different drive and sample voltages, i.e. the $V_{tip} = V_{sample} + V_{drive} \cos(wt)$, is shown in Figure 6 (e, g). From this workflow, we conclude that with a higher AC drive voltage, it requires a smaller sample voltage to flip the domains.

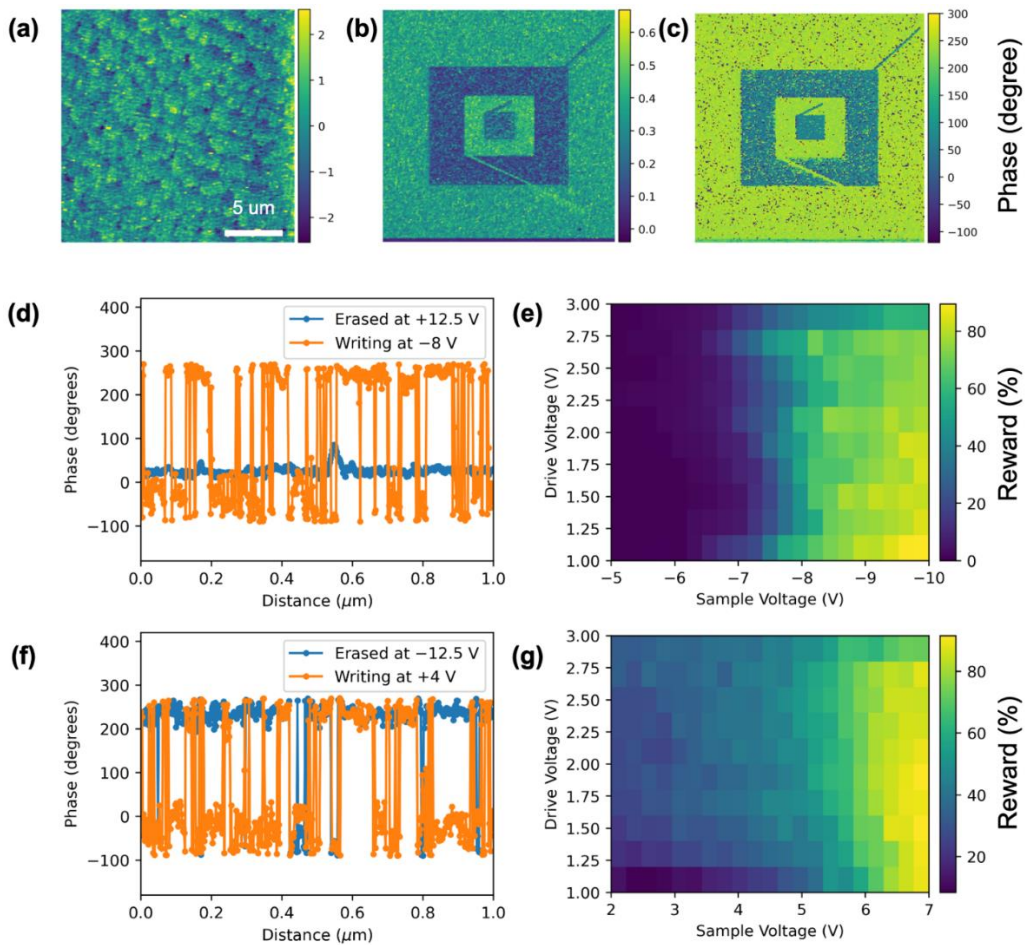

**Figure 5.** Domain writing optimization on a ~150 nm thick $Pb_{0.995}(Zr_{0.45}Ti_{0.55})_{0.99}Nb_{0.01}O_3$ film in the PFM mode. **(a-c).** Domain poling at ±5 V DC tip voltage. (a) height, (b) amplitude and (c) phase channels of the dual amplitude resonance tracking (DART) map taken around a poled pattern. **(d).** The trace lines of the DART phase after erasing

the area with +12.5 V (blue line) and after writing at – 8 V (orange line) of the surface voltage. The reward function is defined as the percent of phase pixels that has flipped from the blue line to the orange line. **(e).** The reward function plotted at different writing sample DC voltage and writing drive voltage. It shows that the required sample DC voltage becomes smaller at larger drive voltage. **(f-g).** Similar grid measurements taken for positive sample writing voltage.

## IV.b. Combinatorial library

As shown in Figure 2, the interface library was used to construct a grid search workflow on a large combinatorial library sample (approximately 10 cm in diameter). In this workflow, the sample was divided into 20 grid points along the direction of composition gradient. At each grid point, the workflow brings the probe to the sample surface, tunes the probe if the imaging mode requires, and takes several topographic scans at different sizes and resolutions. After that, the probe is retracted to a safe distance from the sample surface and the sample moved to the next grid point for the next cycle of measurement.

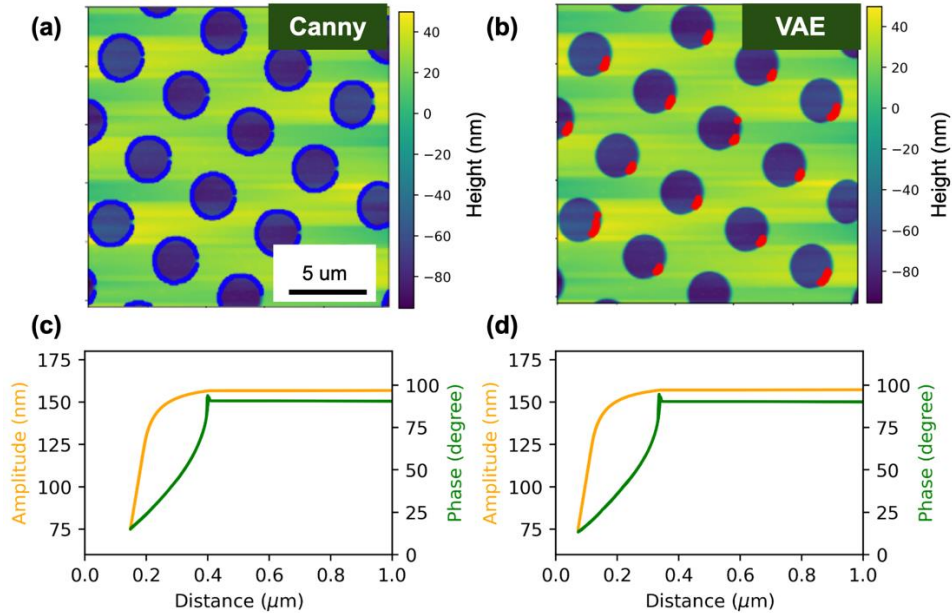

**Figure 6.** Discovery based on structural features. The sample used here is an AR Height calibration grating sample and the patterns are 100 nm deep holes fabricated on a silicon substrate. **(a).** Edge detection by a Canny filter from a local computer. All the edge points around the circular holes are detected and labeled with blue points. **(b).** Structural feature extraction with VAE running on a remote HPC. Compared to the simple Canny filter, VAE offers more control knobs, but requires higher computing power. Only the edge points at the lower right side of the circular holes are selected by VAE and labeled with red markers. **(c).** An amplitude-phase vs z curve measured at an edge location detected by Canny filter in (a). **(d).** Similar amplitude-phase vs z curve measured at an edge location detected by VAE in (b).

## IV.c. Discovery based structural features

In this example, a fixed-policy workflow is shown to take spectroscopy measurements only on selected features. To illustrate this capability, we use the standard grid sample representing circular holes in Si and demonstrate simple feature finding to identify object of interest. In Figure 6(a), the acquired topographic image on a calibration silicon grating sample from Asylum Research (AR) is fed through a Canny filter to detect the presence of edges on a local computer. Then the workflow guides the probe to take force-distance spectroscopy only at these edge points. In Figure 6(b), the acquired topographic image is streamed to the code running on a remote supercomputer (Infrastructure for Scientific Applications and Advanced Computing, or ISAAC). After that, the workflow extracts the structural features in the image into a 2D latent space, from where human operators can subsequently select the edge points only at a specific angle. Finally, the workflow on the ISAAC will move the probe only to these selected locations to take spectroscopy (Figure 6(c-d)). In this example workflow, the policy is fixed upon the detection of edge points and taking force-distance data on them. Note that the force-distance spectroscopy measured on the edges of the silicon grating sample is not intended to provide any physical insights. Instead, it serves as a demonstration of our capability to perform similar tasks, such as I-V curve measurements, hysteresis loop analysis, and force-distance spectroscopy, specifically on domain boundaries or around specific defects as long as these can be identified from topographic, phase, or other imaging channels in SPM.

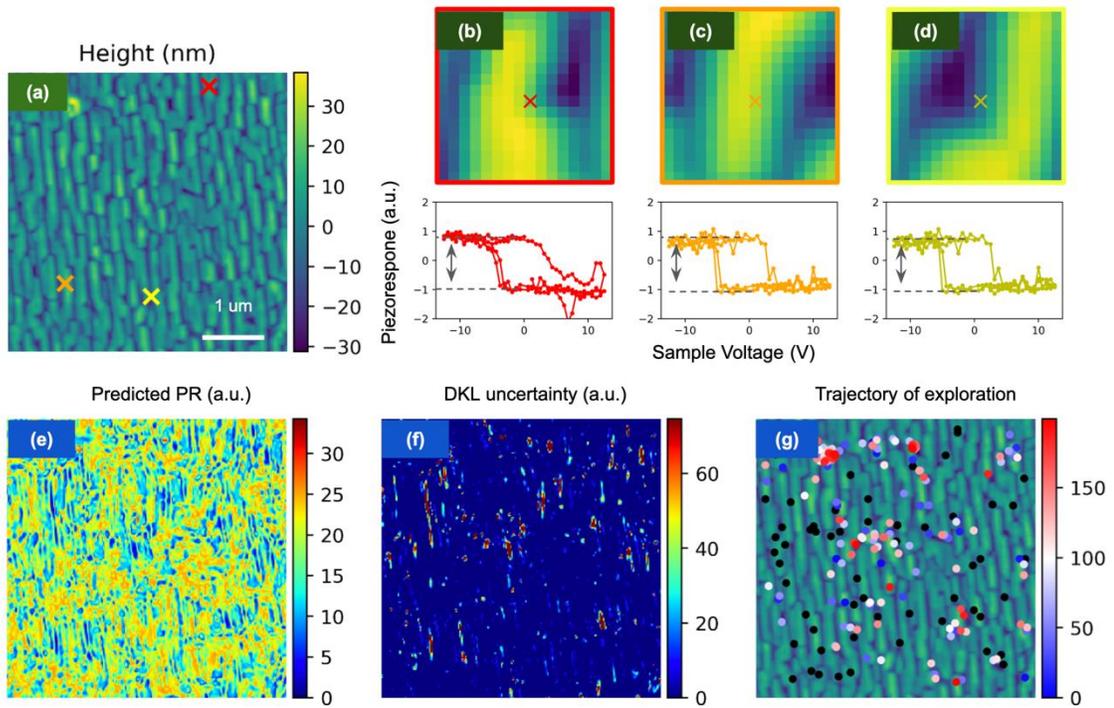

**Figure 7.** Discovery based on spectral features. **(a).** The topographic image taken on ferroelectric $PbTiO_3$. **(b-d).** Example of segmented topographic patches and corresponding hysteresis loop spectra. On selection, the probe is moved to the center of the topographic patch to take a hysteresis loop. Then the loop height will be extracted as the scalarizer. DKL learns the relation between all the measured spectra (scalarizers) and their corresponding topographic patches. The trained model will subsequently be used to predict the distribution of loop height in the whole topographic image in (a). **(e).** Predicted loop height by DKL after 58 steps of seeding plus 200 steps of exploration. **(f).** The DKL uncertainty about its prediction in (e). **(g).** The DKL exploration trajectory shows all the visited points in the seeding (black) and training process (colored).

### IV.d. Deep Kernel Learning

We also reproduced the Deep Kernel Leaning (DKL) workflow from previous work [29]. In this workflow, a topographic scan is first taken on the sample surface. Then the acquired image is streamed to the ISAAC, where the image is segmented into small patches. In the seeding stage, the codes running on ISAAC move the probe to the center of seeding patches and initiate a hysteresis loop spectroscopy measurement, from which the physical property – the maximum piezoresponse (PR) – can be derived as the scalarizer. Here, the scalarizer is a parameter derived from the spectrum that can be correlated with the property of the material. Then the DKL algorithm will train a model to learn the correlation between the measured maximum PR and the structural features in the corresponding patch. Subsequently, this model can be used to predict the distribution of the maximum PR based on the acquired topographic image. In this example, the policy is to search for structural features that can give rise to the maximum PR and the reward function is the PR computed from the measured hysteresis loop. In the active learning stage, the workflow decides where to measure next based on the predicted PR distribution, and it will re-train the model to include the newly acquired spectrum. In the end, we obtain a model that can predict the distribution of PR (Figure 6(d)) based on the topographic image and the accompanying model uncertainty (Figure 6(e)).

### V. From single task to workflow optimization

In the examples above, we have illustrated the integration between the Jupiter AFM and the HPC, the deployment of the workflows that allow grid-based exploration of the image quality as a function of control parameters, image-based spectroscopic measurements, grid based combinatorial library exploration, and ultimately implementation of the deep kernel learning workflow. These are the examples of tasks that are most common in the SPM exploration of real-world systems. These single task workflows with fixed policy or driven by reward functions can be further extended to active learning optimization frameworks.

We further note that the availability of the engineering controls and single task workflows allows a straightforward extension to more complex multi-step discovery workflows that can be driven either by a human operator or organized to pursue reward functions established using human heuristics or large language models. With these, we believe that the connection between the active learning and experimental physical sciences is now possible.

### Methods

### Sample growth

The $(CrVTaW)_xMo_{1-x}$ thin film was grown via dc magnetron co-sputtering from a 50 mm diameter Mo and an equiatomic CrVTaW target at 500°C substrate temperature. The system was pumped to ~ $3x10^{-7}$ Torr and backfilled with Ar to 5 mTorr and the sputtering powers (200 W for CrVTaW and 100 W for Mo) were adjusted to give approximately equivalent sputtering rates (10 nm/min determined via x-ray reflectance) of the two targets at the substrate center. The pseudo binary $(CrVTaW)_xMo_{1-x}$ composition varies from 15 <x <88 at. % across the 100 mm diameter substrate with a roughly linear composition gradient.

The PbTiO$_3$ (PTO) thin films were grown on La$_{0.7}$Sr$_{0.3}$MnO$_3$ (LSMO) buffered (110)-oriented SrTiO$_3$ (STO) substrates using pulsed laser deposition (PLD) with a KrF excimer laser ($\lambda$ = 248 nm). The LSMO/PTO layers were deposited at temperatures of 650 °C/690 °C with oxygen pressures of 90 mtorr/150 mtorr, respectively. After deposition, the samples were cooled to room temperature under an oxygen pressure of 600 Torr. The thicknesses of the PTO and LSMO layers are approximately 150 nm and 30 nm, respectively.

Pb$_{0.995}$(Zr$_{0.45}$Ti$_{0.55}$)$_{0.99}$Nb$_{0.01}$O$_3$ films were grown by pulsed laser deposition using a KrF excimer laser from a ceramic target onto a SrRuO$_3$-electroded (001) SrTiO$_3$ single crystal. The SrRuO$_3$ film was grown from a target from Kojundo Chemical Lab. Co. Ltd., using a laser energy density of 1.5 J/cm$^2$, a substrate temperature of 660°C, an oxygen pressure of 120 mTorr, a target-to-substrate distance of 6.7 mm, and a frequency of 5 Hz. The SrRuO$_3$ film thickness was around 50 nm. The PZT film was grown from a target with 20% excess PbO to compensate for lead loss during growth, using a laser energy density of 1.5 J/cm$^2$, a substrate temperature of 630°C, an oxygen pressure of 120 mTorr, a target-to-substrate distance of 6.2 mm, and a frequency of 5 Hz. The PZT film thickness was around 147 nm.

## Acknowledgements

The development of automated SPM (YL, SVK) was supported by the center for 3D Ferroelectric Microelectronics (3DFeM), an Energy Frontier Research Center funded by the U.S. Department of Energy (DOE), Office of Science, Basic Energy Sciences under Award Number DE-SC0021118. The combinatorial library growth (RE, PDR) was supported by the National Science Foundation Materials Research Science and Engineering Center program through the UT Knoxville Center for Advanced Materials and Manufacturing (DMR-2309083). Growth of PZT samples (SU, STM) was supported by DMR-2025439. PTO (110) growth (JCY, YCL) acknowledged the financial support from National Science and Technology Council (NSTC) in Taiwan, under grant no. NSTC 112-2112-M-006-020-MY3.